\documentclass[
 reprint,
 amsmath,amssymb,
 prl,showkeys
]{revtex4-1}

\usepackage{graphicx}
\usepackage{dcolumn}
\usepackage{bm}
\usepackage{subfigure}
\usepackage{amsmath}
\usepackage{changes}

\begin{document}


\title{Asymmetric access to information impacts the power-law exponent in networks}

\author
{Zhenfeng Cao$^{1}$, Zhou He$^{2,3}$, and Neil F. Johnson$^{1}$}
\affiliation{$^1$Physics Department, University of Miami, Coral Gables, FL 33146, U.S.A.}
\affiliation{$^2$School of Economics and Management, University of Chinese Academy of Sciences, Beijing 100190, China}
\affiliation{$^3$Department of Industrial \& Systems Engineering, National University of Singapore, Singapore 117576}

\date{\today}

\begin{abstract}
The preferential attachment (PA) process is a popular theory for explaining network power-law degree distributions. In PA, the probability that a new vertex adds an edge to an existing vertex depends on the connectivity of the target vertex. In real-world networks, however, each vertex may have asymmetric accessibility to information. Here we address this issue using a new network-generation mechanism that incorporates asymmetric accessibility to upstream and downstream information. We show that this asymmetric information accessibility directly affects the power-law exponent, producing a broad range of values that are consistent with observations. Our findings shed new light on the possible mechanisms in three important real-world networks: a citation network, a hyperlink network, and an online social network.

\end{abstract}


\maketitle

Understanding the mechanisms underlying the power-law degree distribution observed in many complex networks has been a hot research topic for many years \cite{newman2005power, clauset2009power, albert2002statistical}. Preferential attachment (PA) theory has been proposed \cite{jeong2003measuring, albert2002statistical} to explain the growth of diverse systems such as scientific collaboration networks \cite{newman2001clustering, jeong2003measuring, lehmann2005life, milojevic2010modes, kuhn2014inheritance}, the World Wide Web \cite{adamic2000power, faloutsos1999power, barabasi1999emergence, jeong2003measuring, capocci2006preferential}, actor collaboration graphs \cite{barabasi1999emergence, zhang2006model, barabasi2009scale}, social networks \cite{mislove2007measurement, fu2008empirical, liljeros2001web, jones2003assessment, mislove2008growth, ribeiro2010myspace, adamic2003friends, wilson2009user}, and chemical and biological networks \cite{kaiser2004edge, sole2002model, barabasi2004network, eisenberg2003preferential}. In a PA model, the probability that a new vertex will add an edge to an existing vertex depends on the connectivity of the target vertex \cite{barabasi1999emergence}. Although the PA model has many variants \citep[see, e.g.,][]{albert2002statistical, lehmann2005life, choromanski2013scale, pammolli2007generalized, roth2005generalized, dorogovtsev2000structure}, most of them have either one or both of the following drawbacks: (1) Some assumptions and parameters in PA-based models are not well justified. These include the parameters $A$ and $\alpha$ in Ref. \cite{albert2002statistical}, and the ``ghost'' citation in Ref. \cite{lehmann2005life}. (2) The PA rule applies to all vertexes assuming that each vertex has full information about connectivity, which may not be consistent with reality \cite{vazquez2000knowing}. In fact, local network-generation rules can also produce power-law behaviors, as reported by {V{\'a}zquez} \cite{vazquez2003growing,vazquez2000knowing}. However, the effect of asymmetric accessibility to upstream and downstream information remains a largely unexplored issue. \par
In this paper, we present a new network-generation mechanism based on simple local rules that takes into account the asymmetric accessibility to upstream and downstream information. In addition, the physical meanings of all the parameters introduced in our model have clear real-world correspondences, and the power-law exponents of the networks that it produces can be any value greater than 2. We analyze in detail how this asymmetric accessibility affects the power-law exponents. Finally, we show that the proposed network-generation mechanism and findings are consistent with various real-world datasets.\par

\begin{figure}[h]
	\includegraphics[width=\linewidth]{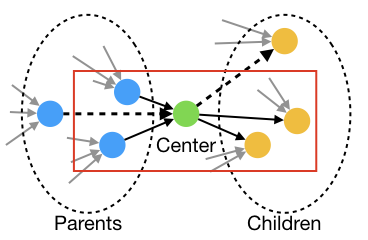}
	\caption{The family of a center node is comprised of nodes in the red solid rectangle: the center (green), its accessible parents (blue), and its accessible children (yellow). The solid arrows (edges) connect nodes whose information is accessible, while the dashed ones connects nodes whose information is inaccessible. The directions of the arrows represent the directions of information flow (e.g., in a citation network, the target node cites the source node).}
	\label{fig1}
\end{figure}

We will use the example of a citation network to introduce our mechanism. Suppose that an author decides to cite some published papers from the network. The author picks one published paper to cite, with that paper effectively being drawn randomly from the network. We label the selected paper as the \textit{center}. We define the articles cited by the center as its \emph{parents}, and the articles citing the center as its \emph{children}. With such definitions, we can see that information flows from parents to children. Due to the finite cost of searching in terms of time spent, it makes sense to assume that the author cannot access the complete set of all parents and children of the center, leading to the situation that the actual information flows in the citation network have an additional asymmetry introduced (e.g., different accessibilities to parents and children). We let $a$ and $b$ be the probability that a parent and a child can be accessed, respectively. A center's \textit{family} is a set comprised of the center, its accessible parents and children, as shown in Fig. \ref{fig1}. Eventually, the author randomly selects one member from the family and cites the selected family member. Consequently, the author's article will be a child of the cited paper. We assume that in each time step, only one article is published and thus it can be uniquely identified by its publication time. In addition, each newly-published article on average will be cited independently $m$ times. \par

Mathematically, let $p(q,s,t)$ denote the probability of an article published at time step $s$ having $q$ children when observed at step $t$. The probability density function (PDF) of the number of children observed at time $t$ can be expressed as $P(q,t)=\frac{1}{t}\sum_{s=0}^{t}p(q,s,t)$. Consider an article  which was published at $s$ and has degree $q$ at time step $t$. The probability that it will be cited by an article published at $t+1$ is given by 
\begin{equation}
A(q, t) = \frac{m}{t} \times \frac{1}{am+bq+1} + \frac{m}{t}\times \frac{aq + bm}{am + b\bar{q}+1},
\end{equation}
where the first term represents the probability that the article will be cited as a center, the second term represents the probability that it will be cited as a parent or a child, and $\bar{q}$ is the average $q$ which equals $m$. If we assume $a q+b m$ is much greater than 1, the first term is then usually much smaller than the second one when $q$ is $\gtrapprox \bar{q}$. This is very likely to be true in real-world datasets: for example,  the average numbers of citations ($m$, or $\bar{q}$) for an article in a citation network, for hyperlinks of a web page on the Internet, and follows for a user in the online social network studied in this work, are found to be  around 10.8, 8.2, and 29.8 respectively. Hence, for simplicity, we replace $q$ in the first term by $\bar{q}$, and have
\begin{equation}
A(q, t) \approx \frac{m}{t} \times \frac{a q + b m + 1}{(a+b)m +1}.
\end{equation}\par
Therefore, we have $p(q,s,t+1)-p(q,s,t)=-A(q,t) p(q,s,t)+A(q-1,t) p(q-1,s,t)$ with the initial condition that $p(q, s, s)=\delta_{q,0}$. Summing over $s$ from $0$ to $t$, and in the long-time limit after transition to continuous-time approximation, we obtain 
\begin{equation}
\begin{split}
\frac{\partial P(q, t)}{\partial t} = &- \frac{P(q, t)}{t} -A(q,t) P(q,t) \\
&+A(q-1,t) P(q-1, t) + \delta_{q, s}\ . 
\end{split}
\end{equation}
Consequently, the solution for the stable state (i.e., $\partial P(q, t)/\partial t=0$ when $t\to \infty$) is given by
\begin{equation}\label{eqn:P}
P(q) = C^{-1} B \left(2+ \frac{b}{a}+\frac{1}{am},\frac{b m+1}{a}+q\right),
\end{equation}
where $B$ is the beta function and $C$ is the normalization constant given by
\begin{equation}
C= \, _2\tilde{F}_1\{1,\frac{b m+1}{a};\frac{2 a+b+b m+1}{a}+\frac{1}{am};1 \} \Gamma\left[\frac{b m+1}{a} \right],
\end{equation}
where $\, _2\tilde{F}_1$ is the hypergeometric function, and $\Gamma$ is the Gamma function. For large $q$, this produces a power-law distribution with the power-law exponent 
\begin{equation}\label{eqn:alpha}
\alpha=2+ \frac{b}{a}+\frac{1}{am}
\end{equation}
which can be any value greater than 2.\par
\begin{figure}[h]
\includegraphics[width=\linewidth]{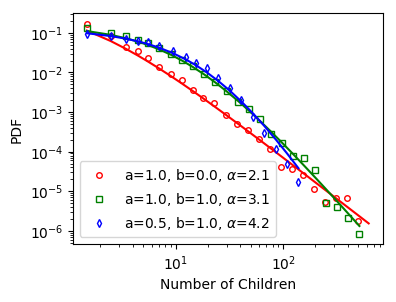}
\caption{Comparison the analytical results (solid lines; i.e., Eq. \ref{eqn:P}) and the simulation results (the symbols of the same color) for different parameter values. $a$ ($b$) is the probabilities that a parent (child) is accessible, and $m$ is the average member of parents a center has.  For all cases here, $m=10$. }
\label{fig2}
\end{figure}

We now present a comparison between these analytical results and simulation in three scenarios, as illustrated in Figure \ref{fig2}. The findings show that: (1) The simulation outputs are very close to the analytical results, which helps validate our model theoretically. (2) The parameter $a$ has a negative relationship with the power-law exponent $\alpha$ meaning that as $a$ decreases, the observed slope in Fig. 2 increases (see green and blue scenarios in Fig. \ref{fig2}). Therefore, as the accessibility to upstream nodes gets reduced, the power-law exponent becomes smaller. It also becomes less likely that a node can have a large number of children. (3) The parameter $b$ and $\alpha$ are positively related (see red and green scenarios in Figure \ref{fig2}). Interestingly, while increasing  $a$ will make $\alpha$ increase linearly (and there is an upper bound for $\alpha$ since $a \leq 1$) when $b$ is given, by decreasing $b$ with $a$ given $\alpha$ will increase acceleratingly and can reach any value above 2.\par

We now compare our model (i.e. Eq. \ref{eqn:P}) to three real-world networks: the citation network of the American Physical Society (available at \emph{https://journals.aps.org/datasets}), the hyperlink network of \emph{www.stanford.edu} \cite{leskovec2009community}, and the relationship network of Twitter \cite{leskovec2012learning}. Let $O_i$ denote the empirical $P(q_i)$ for the $i$th $q$ (i.e., $q_i$) and $E_i$ denotes the theoretical one. Our model is fit to the empirical probability distribution by minimizing $\chi^2$, where
\begin{equation}
\chi^2 = \sum_i \left[\frac{O_i -E_i}{E_i} \right]^2\ .
\end{equation}\par
\begin{figure}[ht]
\includegraphics[width=\linewidth]{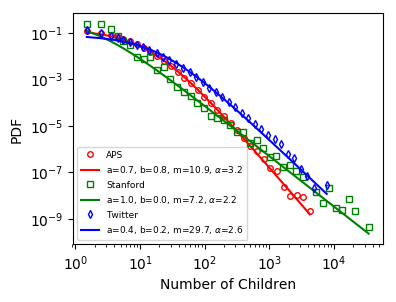}
\caption{Best fit (solid lines) for the three empirical distributions (symbols of same color) using the theoretical model (Eq. \ref{eqn:P}). Optimal fitting parameters are shown in the legend.}
\label{fig3}
\end{figure}

In this way, we deduce the best-fit values of $a$, $b$ and $m$. We find the following: (1) in the APS citation network, the probability that a parent is accessed ($a=0.7$) is very close to the probability that a child is accessed ($b=0.8$). In other words, the parents and children of a central article are almost equally accessible to authors. There could be many interpretations. For instance, when the early-published and newly-published articles are almost equally available, the authors seemingly have almost equal preferences for them. It is also possible that the APS network is setup such that it is easier to obtain information on parents than on children though newly-published articled are favored, and hence the accessibilities to parents and children show little bias. (2) In the hyperlink network, $a=1$ while $b=0$, which means that a new web page cares primarily about the parents of central pages, perhaps because the page creator can easily access the links on the central pages but has very little information about which web pages will direct to the central pages. (3) In the relationship network of Twitter, $a=0.4>b=0.2$, probably indicating that when adding a new follow, users favor the follows of the current follows over the followers of them. (4) The predicted $m$ for all the three cases (i.e., 10.9, 7.2, 29.7; see Fig. \ref{fig3}) are approximately the same as their corresponding empirical one (i.e., 10.5, 8.2, and 29.8 in turn), which further validates our model empirically. As a result of these findings, we believe that the proposed network-generation mechanism incorporating the effect of asymmetric accessibility to upstream and downstream information, provides a surprisingly simple explanation for the power-law distributions observed in diverse real-world networks.\par
In conclusion, we have presented a new model based on information accessibility that can explain the power-law degree distributions observed in a variety of different real-world networks. Compared with the original PA model, our network-generation mechanism is comprised of local rules which consider the practical situation that the accessibility to upstream and downstream information is asymmetric. We have also investigated the relationships between the asymmetric accessibility to upstream and downstream information and the power-law exponent of the yielded network. We validated our model by exploring the fit to three real-world networks, and demonstrated the empirical impact of asymmetric accessibility to upstream and downstream information on the power-law exponent. We suggest that both theoretical and empirical findings could be insightful for network managers to understand and control real-work networks by adjusting the information accessibilities. 

\bibliography{PA.bib}
\end{document}